\documentclass{svproc}
\usepackage{url}

\usepackage{graphicx}
\usepackage{amsfonts}
\usepackage{hyperref}
\usepackage{enumitem}
\usepackage{bbding}
\usepackage{amsmath}

\begin{document}
\mainmatter              
\title{Scalability Evaluation of NSLP Algorithm for Solving Non-Stationary Linear Programming Problems on Cluster Computing Systems}
\titlerunning{Scalability Evaluation of NSLP Algorithm}
\author{{Irina Sokolinskaya \and Leonid B. Sokolinsky\Envelope}\thanks{The
reported study has been partially supported by the RFBR according to research
project \mbox{No.~17-07-00352-a}, by the Government of the Russian Federation
according to Act 211  (contract \mbox{No.~02.A03.21.0011.}) and by the
Ministry of Education and Science of the Russian Federation (government
order 1.9624.2017/7.8).}}
\authorrunning{I. Sokolinskaya, L.B. Sokolinsky} 
%
\tocauthor{Irina Sokolinskaya and Leonid B. Sokolinsky}
\institute{South Ural State University \\76 Lenin prospekt, Chelyabinsk, Russia, 454080\\
\email{Irina.Sokolinskaya@susu.ru}, \email{Leonid.Sokolinsky@susu.ru}}

\maketitle              

\begin{abstract}
The paper is devoted to a scalability study of the NSLP algorithm for solving non-stationary high-dimension linear programming problem on the cluster computing systems. The analysis is based on the BSF model of parallel computations. The BSF model is a new parallel computation model designed on the basis of BSP and SPMD models. The brief descriptions of the NSLP algorithm and the BSF model are given. The NSLP algorithm implementation in the form of a BSF program is considered. On the basis of the BSF cost metric, the upper bound of the NSLP algorithm scalability is derived and its parallel efficiency is estimated. NSLP algorithm implementation using BSF skeleton is described. A comparison of scalability estimations obtained analytically and experimentally is provided.

\end{abstract}
\section{Introduction}
The Big Data phenomenon has spawned the large-scale linear programming (LP) problems~\cite{sokol_Chung}. Such problems arise in the following areas: scheduling, logistics, advertising, retail, e\nobreakdash-\hspace{0pt}commerce~\cite{sokol_Tipi}, quantum physics~\cite{sokol_Gondzio}, asset-liability management~\cite{sokol_Sodhi}, algorithmic trading~\cite{Brogaard,Budish,Gomber,Hendershott} and others. The similar LP problems include up to tens of millions of constraints and up to hundreds of millions of decision variables. In many cases, especially in mathematical economy, these LP problems are nonstationary (dynamic). It means that input data (matrix $A$, vectors $b$ and $c$) is evolving with time, and the period of data change is within the range of hundredths of a second.

Until now, one of the most popular methods solving LP problems is the class of algorithms proposed and designed by Dantzig on the base of the simplex method~\cite{Dantzig}. The simplex method has proved to be effective in solving a large class of LP problems. However, in certain cases the simplex method has to move across all the vertices of the polytope, which corresponds to an exponential time complexity~\cite{Klee}. Karmarkar in~\cite{Karmarkar} proposed a method for linear programming called ``Interior point method'' which runs in polynomial time and is also very efficient in practice.

The simplex method and the method of interior points remain today the main methods for solving the LP problem. However, these methods may prove ineffective in the case of large scale LP problems with rapidly evolving input data. To overcome the problem of non-stationarity of input data, the authors proposed in~\cite{Sokolinskaya2017} the scalable algorithm \emph{NSLP} (\emph{Non-Stationary Linear Programming}) for solving large-scale non-stationary LP problems on cluster computing systems. It includes two phases: \emph{Quest} and \emph{Targeting}.
The \emph{Quest} phase calculates a solution of the system of inequalities defining the constraint system of the linear programming problem under condition of the dynamic changes of input data. The point of pseudo-projection on $n$\nobreakdash-\hspace{0pt}polytope $M$ is taken as a solution. Polytope $M$ is the set of feasible solutions of the LP problem. The pseudo-projection is an extension of the projection, which uses Fejer (relaxation) iterative process~\cite{Agmon,Motzkin,Eremin,González}. A distinctive feature of the Fejer process is its "self-guided" capability: the Fejer process automatically corrects its motion path according to the polytope position changes during the calculation of the pseudo-projection. The \emph{Quest} phase was investigated in~\cite{Sokolinskaya2017}, where the convergence theorem was proved for the case when the polytope is translated with a fixed vector in the each unit of time. In the paper~\cite{Sokolinskaya2016}, the authors demonstrated that Intel Xeon Phi multi-core processors can be efficiently used for calculating the pseudo-projections.

The \emph{Targeting} phase forms a special system of points having the shape of the $n$\nobreakdash-\hspace{0pt}dimensional axisymmetric cross. The cross moves in the $n$\nobreakdash-\hspace{0pt}dimensional space in such a way that the solution of the LP problem permanently was in the $\varepsilon$\nobreakdash-\hspace{0pt}vicinity of the central point of the cross. The Targeting phase can be effectively implemented as a parallel program for a clustered computing system by using the "master-workers" framework \cite{Sahni,Silva,Leung}. In this paper, we discuss a parallel implementation of the NSLP algorithm using the BSF computational model presented in \cite{Sokolinsky}. On the base of the described BSF\nobreakdash-\hspace{0pt}implementation, a quantitative scalability analysis of the NSLP algorithm is performed.

The rest of the paper is organized as follows. Section 2 gives a formal statement of a LP problem and presents the brief description of the NSLP algorithm. Section 3 provides an outline of the BSF computational model and presents corresponding cost metrics. Section 4 describes a BSF\nobreakdash-\hspace{0pt}implementation of the NSLP algorithm, calculates the upper bound of scalability and evaluates the parallel efficiency depending on the percentage of initial data being changed dynamically. Section 5 describes an implementation of the NSLP algorithm based on the BSF skeleton in C language and compares the results obtained analytically and experimentally. Section 6 summarizes the results obtained and proposes the directions for future research.

\section{NSLP algorithm} \label{NSLP_algorithm}

Let we be given a non-stationary LP problem in the vector space ${\mathbb{R}^n}$:
\begin{equation} \label{Eq1}
{\text{max}}\left\{{\left\langle {{c_t},x} \right\rangle |{A_t}x \leq {b_t},\;x \geq 0} \right\},
\end{equation}
where the matrix ${A_t}$ has $m$ rows. The non-stationarity of the problem means that the values of the elements of the matrix ${A_t}$ and the vectors ${b_t}$, ${c_t}$ depend on the time $t \in {\mathbb{R}_{ \geq 0}}$. We assume that the value of $t = 0$ corresponds to the initial instant of time:
\begin{equation} \label{Eq2}
{A_0} = A, {b_0} = b, {c_0} = c
\end{equation}

\begin{figure}
  \centering
  \includegraphics[scale=0.9]{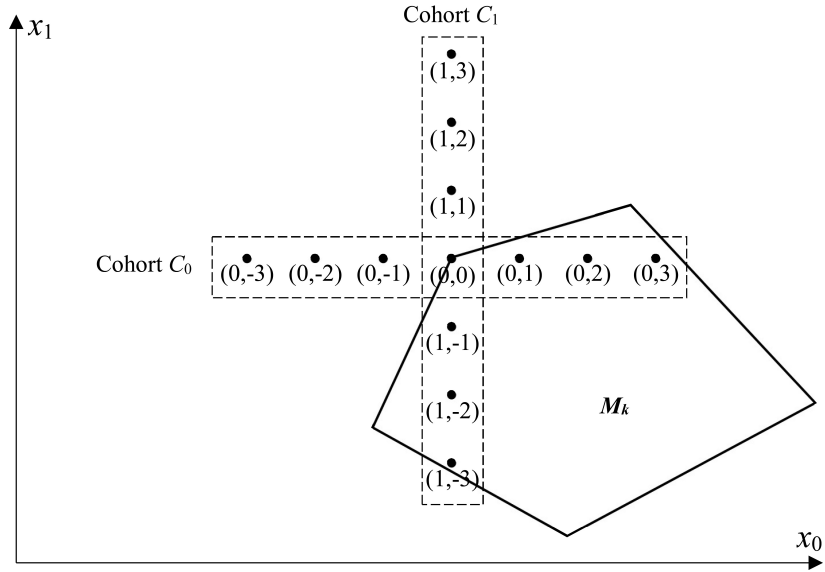}
  \caption{Two-dimensional cross $G$: $K = 6$, $P = 12$}
  \label{Fig1}
\end{figure}

Let ${M_t}$ be a polytope defined by the constraints of the non-stationary LP problem (\ref{Eq1}). Such a polytope is always convex. The \emph{Quest} phase calculates a point $z$ belonging to the polytope ${M_t}$. This phase is described in detail in~\cite{Sokolinskaya2017}. The \emph{Quest} Phase is followed by the \emph{Targeting} phase. At the \emph{Targeting} phase, a $n$\nobreakdash-\hspace{0pt}dimensional axisymmetric cross is formed. The \emph{n\nobreakdash-\hspace{0pt}dimensional axisymmetric cross} is a finite set $G = \{ {g_0}, \ldots ,{g_{P - 1}}\}  \subset {\mathbb{R}^n}$ having the cardinality equals $P + 1$, where $P$ is a multiple of $n \geq 2$. Among points of the cross, the point ${g_0}$ called the \emph{central point} is single out. The initial coordinates of the central point are assigned the coordinates of the point $z$ calculated in the \emph{Quest} phase. The set $G\backslash \{ {g_0}\} $ is divided into $n$ disjoint subsets ${C_i}$ ($i = 0, \ldots ,n - 1$) called the \emph{cohorts}:
\[G\backslash \{ {g_0}\}  = \bigcup\limits_{i = 0}^{n - 1} {{C_i}} .\]
Each $i$\nobreakdash-\hspace{0pt}th cohort ($i = 0, \ldots , n - 1$) consists of
\begin{equation} \label{Eq3}
K = P/n
\end{equation}
points lying on the straight line, which is parallel to the $i$\nobreakdash-\hspace{0pt}th coordinate axis and passing through the central point ${g_0}$. By itself, the central point does not belong to any cohort. The distance between any two neighbor points of the set $G \cup \{ {g_0}\}$ is equal to the constant~$s$. It can be changed during computing. An example of the two-dimensional cross is shown in Fig.~\ref{Fig1}. The number of points in one dimension excluding the central point is equal to $K$. The symmetry of the cross supposes that $K$ takes only even values greater than or equal to~2. Using equation~(\ref{Eq3}), we obtain the following equation giving the total number of points in the cross:
\begin{equation} \label{Eq4}
P + 1 = nK + 1
\end{equation}
Since $K$ can take only even values greater than or equal to 2 and $n \geq 2$, from equation~(\ref{Eq4}), it follows that $P$ can also take only even values and $P \geq 4$. In Fig.~\ref{Fig1}, we have $n = 2$, $K = 6$, $P = 12$.

Each point of the cross $G$ is uniquely identified by a \emph{marker} being a pair of integers numbers $(\chi ,\eta )$ such that $0 \leq \chi  < n$, $\left| \eta  \right| \leq K/2$. Informally, $\chi $ specifies the number of the cohort, and $\eta $ specifies the sequence number of the point in the cohort ${C_\chi }$, being counted out of the central point. The corresponding marking of points for the two-dimensional case is given in Fig.~\ref{Fig1}. The coordinates of the point ${x_{(\chi ,\eta )}}$ having the marker $(\chi ,\eta )$ can be reconstructed as follows:
\begin{equation} \label{Eq5}
{x_{(\chi ,\eta )}} = {g_0} + (0, \ldots ,0,\underbrace {\eta  \cdot s}_\chi ,0, \ldots ,0)
\end{equation}
The vector being added to ${g_0}$ in the right part of the equation~(\ref{Eq5}) has a single non-zero coordinate in the position $\chi $. This coordinate equals $\eta \cdot s$, where $s$ is the distance between neighbor points in a cohort.

The \emph{Targeting} phase includes the following steps.
\begin{enumerate}
\item Build the $n$\nobreakdash-\hspace{0pt}dimensional axisymmetric cross
    $G$ that has $K$ points in each cohort, the distance between neighbor
    points equaling $s$, and the center at point ${g_0} = {z_k}$, where
    ${z_k}$ is obtained in the \emph{Quest} phase.

\item Calculate $G' = G \cap {M_k}$.

\item Calculate ${C'_\chi } = {C_\chi } \cap G'$ for $\chi  = 0, \ldots ,n
    - 1$.

\item Calculate $Q = \bigcup\limits_{\chi  = 0}^{n - 1} {\left\{ {\arg \max
    \left\{ {\left\langle {{c_k},g} \right\rangle \mid {g \in {{C'}_\chi
    }},{{C'}_\chi } \ne \emptyset } \right\}} \right\}} $.

\item If ${g_0} \in {M_k}$ and $\left\langle {{c_k},{g_0}} \right\rangle
    \ge \mathop {\max }\limits_{q \in Q} \left\langle {{c_k},q}
    \right\rangle $, then $k: = k + 1$, and go to step~2.

\item ${g_0}: = \frac{{\sum\limits_{q \in Q} q }}{{\left| Q \right|}}$.

\item $k: = k + 1$.

\item Go to step~2.
\end{enumerate}

Thus, in the Targeting phase, the steps 2\nobreakdash--7 form a perpetual loop in which the approximate solution of the non-stationary LP problem is permanently recalculated. From the non-formal point of view, in the step 2, we determine which points of the cross $G$ are belonged to the polytope ${M_k}$. In the step 3, points that do not belong to the polytope are dropped out of each cohort. In the step 4, the point with the maximum value of the objective function is chosen among the residuary points of each cohort. In the step 5, we check if the value of the objective function at the central point of the cross is greater than all the maximums found in the step 4. If this condition is true then the cross does not shift, the time counter t is incremented by one unit and the next iteration is started. If this condition is false then we go to step~6 where the new center point is calculated as the centroid of the set of points obtained in the step 4. In the step 7, the time counter t is incremented by one unit. In the step 8, we go to the new iteration. In such a way, the center ${g_0}$ of the cross $G$ permanently performs the role of an approximate solution of the non-stationary problem~(\ref{Eq1}).

\section{BSF computational model}\label{BSF_model}

We use the BSF parallel computation model proposed in \cite{Sokolinsky} to evaluate the upper bound of the scalability of the NSLP algorithm in the Targeting phase. The BSF (Bulk Synchronous Farm) model was proposed to multiprocessor systems with distributed memory. A \emph{BSF-computer} consists of a collection of homogeneous computing nodes with private memory connected by a communication network that delivers messages among the nodes. Among all the computing nodes, one node called the \emph{master-node} is single out. The rest of the nodes are the \emph{slave-nodes}. The BSF-computer must include at least one master-node and one slave-node. Thus, if $P$ is the number of slave-nodes then $P \geq 1$.

\begin{figure}
  \centering
  \includegraphics[scale=0.9]{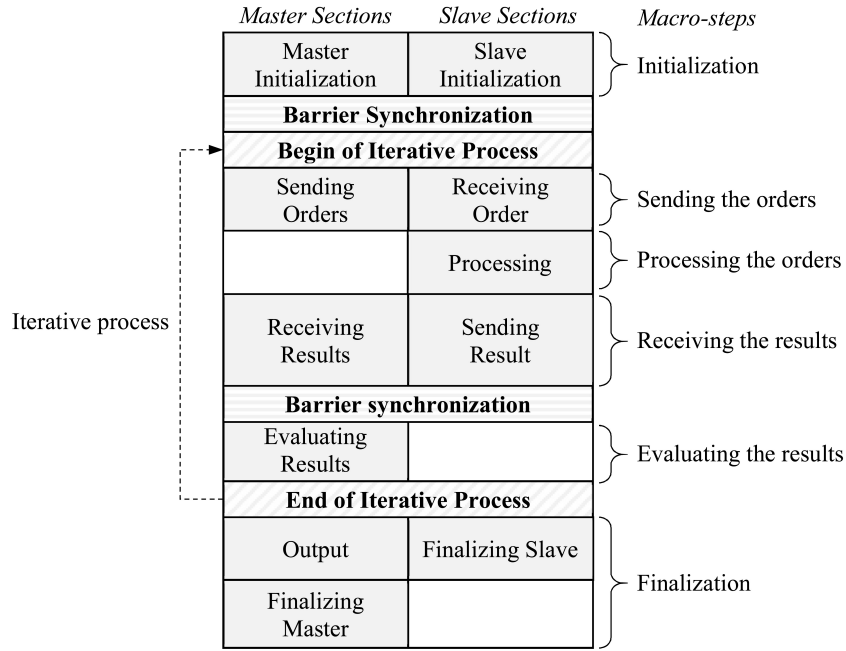}
  \caption{BSF-program structure}
  \label{Fig2}
\end{figure}

BSF-computer utilizes the \emph{SPMD} programming model \cite{Darema} according to which all nodes executes the same program but process different data. A BSF-program consists of sequences of macro-steps and global barrier synchronizations performed by the master and all the slaves. Each macro-step is divided into two sections: master section and slave section. A master section includes instructions performed by the master only. A slave section includes instructions performed by the slaves only. The sequential order of the master section and the slave section within the macro-step is not important. All the slave nodes act on the same data array, but the base address of the data assigned to the slave-node for processing is determined by the logical number of this node. The BSF-program includes the following sequential sections (see Fig.~\ref{Fig2}):
\begin{itemize}
  \item initialization;
  \item iterative process;
  \item finalization.
\end{itemize}

\emph{Initialization} is a macro-step, during which the master and slaves read or generate input data. The initialization is followed by a barrier synchronization. The \emph{iterative process} repeatedly performs its body until the exit condition checked by the master becomes true. In the \emph{finalization macro-step}, the master outputs the results and ends the program.

\emph{Body of the iterative process} includes the following macro-steps:
\begin{enumerate}[label=\arabic*)]
  \item sending the orders (from master to slaves);
  \item processing the orders (slaves);
  \item receiving the results (from slaves to master);
  \item evaluating the results (master).
\end{enumerate}

In the first macro-step, the master sends the same orders to all the slaves. Then, the slaves execute the received orders (the master is idle at that time). All the slaves execute the same program code but act on the different data with the base address depending on the slave-node number.

It means that all slaves spend the same time for calculating. During processing the order, there are no data transfers between nodes. In the third step, all slaves send the results to the master. After that, global barrier synchronization is performed. During the fourth step, the master evaluates received results. The slaves are idle at that time. After result evaluations, the master checks the exit condition. If the exit condition is true then iterative process is finished, otherwise the iterative process is continued.

The BSF model provides an analytical estimation of the scalability of a BSF-program. The main parameters of the model are~\cite{Sokolinsky}:
\begin{itemize}
  \item [$P$:] the number of slave-nodes;
  \item [$L$:] an upper bound on the latency, or delay, incurred in communicating a message containing one byte from its source node to its target node;
  \item [${{t}_{s}}$:] time that the master-node is engaged in sending one order to one slave-node excluding the latency;
  \item [${{t}_{v}}$:] time that a slave-node is engaged in execution an order within one iteration (BSF\nobreakdash-\hspace{0pt}model assumes that this time is the same for all the slave-nodes and it is a constant within the iterative process;
  \item [${{t}_{r}}$:] total time that the master-node is engaged in receiving the results from all the slave-nodes excluding the latency;
  \item [${{t}_{p}}$:] total time that the master-node is engaged in evaluating the results received from all the slave-nodes.
\end{itemize}
Let’s denote ${t_w} = P \cdot {t_v}$ -- summarized time which is spent by slave-nodes for order executions. Then, the upper bound of a BSF-program scalability can be estimated by the following inequality~\cite{Sokolinsky}:
\begin{equation} \label{Eq6}
P \leq \sqrt {\frac{{{t_w}}}{{2L + {t_s}}}}.
\end{equation}
Note that the upper bound of the BSF-program scalability does not depend on the time, which the master is engaged in receiving and evaluating the slave results. The speedup of BSF-program can be calculated by the following equation~\cite{Sokolinsky}:
\begin{equation} \label{Eq7}
a = \frac{{P(2L + {t_s} + {t_r} + {t_p} + {t_w})}}{{{P^2}(2L + {t_s}) + P({t_r} + {t_p}) + {t_w}}}.
\end{equation}
One more important property of a parallel program is the parallel efficiency. The parallel efficiency of a BSF-program can be calculated by the following approximate equation~\cite{Sokolinsky}:
\begin{equation} \label{Eq8}
e \approx \frac{1}{{{{1 + \left( {{P^2}(2L + {t_s}) + P({t_r} + {t_p})} \right)} \mathord{\left/
 {\vphantom {{1 + \left( {{P^2}(2L + {t_s}) + P({t_r} + {t_p})} \right)} {{t_w}}}} \right.
 \kern-\nulldelimiterspace} {{t_w}}}}}.
\end{equation}

\section{BSF-implementation of NSLP algorithm}

In this section, we demonstrate how the algorithm presented in section 2 can be implemented in the BSF-program form. Based on this implementation, we calculate the time complexity of one iteration and give an analytical estimation of the scalability upper bound of the NSLP algorithm in the Targeting phase. For calculating, we use the synthetic scalable linear programming problem of the dimension   called \emph{Model-n}~\cite{Sokolinskaya2016}. This LP problem has the matrix $A$ of the size $n \times 2(n + 1)$. We assume $n > {10^4}$.

The NSLP algorithm can be implemented in the BSF-program form by the following way. In the Initialization macro-step, the master and all the slaves read (generate) and store in the local memory all the initial data of the non-stationary LP problem~(\ref{Eq1}); the master executes the Quest phase and finds a point $z$ belonging to the polytope ${M_t}$. Then, the iterative process begins. In each iteration, the following steps are performed:
\begin{enumerate}[label=\arabic*)]
  \item sending the orders from master to slaves;
  \item processing the orders by slaves;
  \item sending the results from slaves to master;
  \item evaluating the results by master.
\end{enumerate}

\begin{table}[hb!]
\caption{Structure of message ``Order for slaves''}\label{Table1}
\begin{center}
\begin{tabular}{|c|c|p{5cm}|c|}
\hline
\textbf{No.} & \textbf{Attribute ID} & \textbf{Attribute semantic} & \textbf{Overhead} \\
\hline
1 & $\theta$ & New central point coordinates of the $n$\nobreakdash-\hspace{0pt}dimensional cross & ${t_\theta }$ \\
\hline
2 & $\alpha$ & New values of matrix $A$ entries & ${t_\alpha }$\\
\hline
3 & $\beta$ & New values of column $b$ elements & ${t_\beta }$\\
\hline
4 & $\gamma$ & New values of objective function coefficients & ${t_\gamma }$\\
\hline
\end{tabular}
\end{center}
\end{table}

The \emph{order} includes the information given in Table~\ref{Table1}. Suppose that the fraction of the changed elements of matrix $A$, column $b$ and objective function $c$ coefficients equals to $\delta (n)$, where $\forall n\left( {0 \leq \delta (n) \leq 1} \right)$. In that case, the time ${t_s}$ that the master-node is engaged in sending one order to one slave-node (excluding the latency) can be approximated according to Table 1 as follows:
\begin{multline*}
{t_s} = {t_\theta } + {t_\alpha } + {t_\beta } + {t_\gamma } = O(n) + O(\delta (n) \cdot n \cdot 2(n + 1)) + O(\delta (n) \cdot 2(n + 1)) + O(\delta (n)n) = \\
= O(n) + O(\delta (n) \cdot n(n + 1)) + O(\delta (n)(n + 1)) + O(\delta (n)n) < \\
   < O(n + 1) + O(\delta (n) \cdot {(n + 1)^2}) + O(\delta (n)(n + 1)) + O(\delta (n)(n + 1)) =  \\
= O(n + 1) + O(\delta (n) \cdot {(n + 1)^2}) + O(\delta (n)(n + 1)) < \\
< O(\delta (n) \cdot {(n + 1)^2}) + O(n + 1).
\end{multline*}
Hence,
\begin{equation} \label{Eq9}
{t_s} < O(\delta (n) \cdot {(n + 1)^2}) + O(n + 1).
\end{equation}

The smallest unit of parallelization in the BSF-implementation of the NSLP algorithm is the cohort. The number of cohorts equals to the space dimension $n$. Thus, the number $P$ of slave-nodes should be less than or equal to the space dimension $n$. We shall assume $n \gg P$. A slave-node sequentially process all the cohorts assigned to it. In the current cohort, the coordinates of every point $x$ are calculated using equation~(\ref{Eq5}). The time complexity of this operation is $O(n)$. Then the point $x$ is checked to be belonged to the polytope ${M_t}$. To do this, it is sufficient for the slave to verify the truth of the condition ${A_t}x = {b_t}$. Since ${A_t}$ is of size $n \times 2(n + 1)$, the time complexity of this operation is $O({n^2} + n)$. The number of points in a cohort excluding the central one is equal to the constant $K$. According to the equation~(\ref{Eq4}), the total number of points in the cross excluding the central one is equal to $nK$. Hence, the time complexity of the calculations performed for all points of the cross in steps 2\nobreakdash--3 can be estimated as $O({n^3} + {n^2})$. After this, the slaves partially (for their cohorts only) execute the step 4 of Targeting phase (see Section~2). The total time complexity of these operations is $O({n^2})$. Thus, the total time complexity of all the calculations performed by slaves has the following estimation:
\begin{equation} \label{Eq10}
{t_w} = O({n^3} + {n^2}) + O({n^2}) + O(n) \leq O({n^3} + {n^2} + n).
\end{equation}

As a result, each slave sends to the master a summarized vector of points which belong to the polytope and have the maximum value of the objective function in the corresponding cohort. Thus, the total time complexity of transferring the results from the slaves to the master is
\begin{equation} \label{Eq11}
{t_r} = O(PKn) \approx O(n).
\end{equation}

Having received the results from the slaves, the master sums them up to complete the step 4 of the algorithm. The time complexity of these calculations will have the following estimation:
\begin{equation} \label{Eq12}
{t_{{\text{step 4}}}} \approx O({n^2}).
\end{equation}
Because of the non-stationarity of the LP problem, the condition in step 5 will be rarely true. Hence, we may assume that the next step after the step 4 will be the step 6 in most cases. Since the number of cohorts equals to $n$, the total time complexity of the step 6 of the Targeting phase is
\begin{equation} \label{Eq13}
{t_{{\text{step 6}}}} = O({n^2}).
\end{equation}
Thus, the total time complexity of processing the results obtained by the master from the slaves is
\begin{equation} \label{Eq14}
{t_p} \approx {t_{{\text{step 4}}}} + {t_{{\text{step 6}}}} = O({n^2}).
\end{equation}
Substituting the values from~(\ref{Eq10}) and~(\ref{Eq9}) into equation~(\ref{Eq6}), we obtain the following estimation for the upper bound of the NSLP algorithm scalability:
\begin{equation} \label{Eq15}
{P_{NSLP}} \leq \sqrt {\frac{{O({n^3} + {n^2} + n)}}{{2L + O(\delta (n) \cdot {{(n + 1)}^2}) + O(n + 1)}}}.
\end{equation}

Suppose all the input data of the problem are changed at each iteration. It corresponds to $\delta (n) = 1$. In this case, inequality~(\ref{Eq15}) is converted to the following form
\begin{equation} \label{Eq16}
{P_{NSLP}} \leq \sqrt {\frac{{O({n^3} + {n^2} + n)}}{{2L + O({{(n + 1)}^2}) + O(n + 1)}}}  \approx O(\sqrt n) .
\end{equation}
It means that the upper bound of the BSF-program scalability increases proportionally to the square root of the problem dimension. Hence, the NSLP algorithm implementation in the form of a BSF-program has limited scalability in this case.

Now suppose that the fraction of the changed problem input data at each iteration is
\begin{equation} \label{Eq17}
\delta (n) = \frac{1}{{2(n + 1)}}.
\end{equation}
It corresponds to a situation where the matrix   has only one changed row, the column   has only one changed element, and the objective function has no more than one changed coefficient. In this case, we obtain the following estimation
\begin{equation} \label{Eq18}
{P_{NSLP}} \leq \sqrt {\frac{{O({n^3} + {n^2} + n)}}{{2L + O(n + 1) + O(n + 1)}}}  \approx O(\sqrt {n^2})  = O(n)
\end{equation}
substituting the value of $\delta (n)$ from the equation~(\ref{Eq17}) into the equation~(\ref{Eq15}). It means that the upper bound of the BSF-program scalability increases proportionally to the problem dimension. Hence, the NSLP algorithm implementation in the form of a BSF-program is scalable well in this case.

We can also estimate the BSF-implementation parallel efficiency of the NSLP algorithm using approximate equation~(\ref{Eq8}):
\begin{equation} \label{Eq19}
\begin{gathered}
  e = \frac{1}{{1 + \frac{{{P^2} \cdot (2L + {t_s}) + P \cdot ({t_r} + {t_p})}}{{{t_w}}}}} =  \\
   = \frac{1}{{1 + \frac{{{P^2} \cdot (2L + O(\delta (n) \cdot {{(n + 1)}^2}) + O(n + 1)) + P \cdot (O(n) + O({n^2}))}}{{O({n^3} + {n^2})}}}}.
\end{gathered}
\end{equation}
For $\delta (n) = 1$, $n \to \infty $ and $P \to \infty $, we get from~(\ref{Eq19}) the following estimation
\begin{equation} \label{Eq20}
\begin{gathered}
  e = \frac{1}{{1 + \frac{{{P^2}(2L + O({{(n + 1)}^2}) + O(n + 1)) + P(O(n) + O({n^2}))}}{{O({n^3} + {n^2})}}}} \approx  \hfill \\
     \approx \frac{1}{{1 + \frac{{{P^2}(O({n^2}) + O(n)) + P \cdot (O({n^2}) + O(n))}}{{O({n^3} + {n^2})}}}} =  \hfill \\
    = \frac{1}{{1 + ({P^2} + P)\frac{{O({n^2}) + O(n)}}{{O({n^3} + {n^2})}}}} \approx \frac{1}{{1 + \frac{{{P^2} + P}}{{O(n)}}}} \approx \frac{1}{{1 + {P^2}/O(n)}}. \hfill \\
\end{gathered}
\end{equation}
In such a way, we obtain
\begin{equation} \label{Eq21}
e \approx \frac{1}{{1 + {{{P^2}} \mathord{\left/
 {\vphantom {{{P^2}} {O(n)}}} \right.
 \kern-\nulldelimiterspace} {O(n)}}}}.
\end{equation}
Hence for $\delta (n) = 1$, the high parallel efficiency is achieved when $n \gg {P^2}$.

For $\delta (n) = \frac{1}{{2(n + 1)}}$ we get from~(\ref{Eq19}) the following estimation
\begin{equation} \label{Eq22}
e \approx \frac{1}{{1 + {{{P^2}} \mathord{\left/
 {\vphantom {{{P^2}} {O({n^2})}}} \right.
 \kern-\nulldelimiterspace} {O({n^2})}} + {P \mathord{\left/
 {\vphantom {P {O(n)}}} \right.
 \kern-\nulldelimiterspace} {O(n)}}}}.
\end{equation}
Hence for $\delta (n) = \frac{1}{{2(n + 1)}}$, the high parallel efficiency is achieved when $n \gg P$.

\begin{figure}
  \centering
  \includegraphics[scale=0.77]{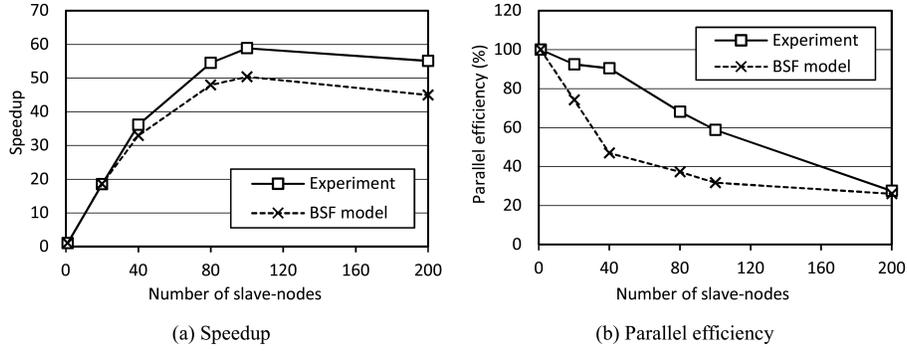}
  \caption{Experiments for $n = 400$.}
  \label{Fig3}
\end{figure}
\begin{figure}
  \centering
  \includegraphics[scale=0.77]{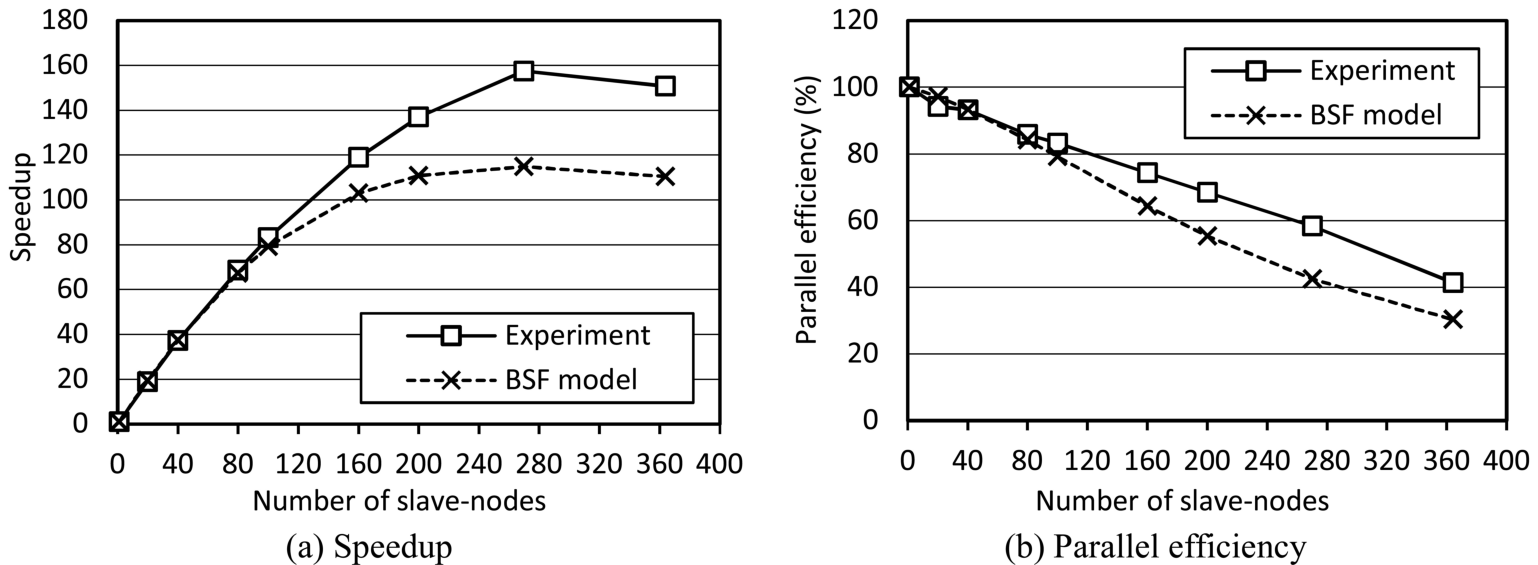}
  \caption{Experiments for $n = 800$.}
  \label{Fig4}
\end{figure}
\begin{figure}
  \centering
  \includegraphics[scale=0.77]{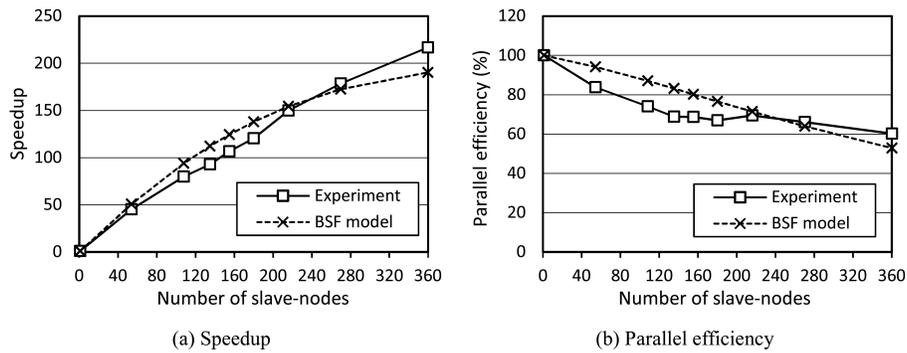}
  \caption{Experiments for $n = 1080$.}
  \label{Fig5}
\end{figure}
\section{Numerical Experiments}

The implementation of the Qwest phase was described and evaluated by us in the paper~\cite{Sokolinskaya2016}. In the present work, we have done the implementation of the Targeting phase in C language using the BSF skeleton. The source code of this program is freely available on Github, at \url{https://github.com/leonid-sokolinsky/BSF-NSLP}. We investigated the speedup and parallel efficiency of this BSF-program on the supercomputer "Tornado SUSU"~\cite{Kostenetskiy} using the synthetic scalable linear programming problem \emph{Model n}~\cite{Sokolinskaya2016} mentioned in the section 4. The calculations were performed for the dimensions 400, 800 and 1080. At the same time, we plotted the curves of speedup and parallel efficiency for these dimensions using equations~(\ref{Eq7}) and~(\ref{Eq8}). We assumed that $\delta (n) = {1 \mathord{\left/  {\vphantom {1 {2(n + 1)}}} \right. \kern-\nulldelimiterspace} {2(n + 1)}}$. The results are presented in Fig.~\ref{Fig3}--\ref{Fig5}. In all cases, the analytical estimations were very close to experimental ones. Moreover, the performed experiments show that the upper bound of the BSF-program scalability increases proportionally to the problem dimension. It was analytically predicted using the equation~(\ref{Eq18}) in Section~4.

\section{Conclusion}

In this paper, the scalability and parallel efficiency of the NSLP algorithm used to solve large-scale non-stationary linear programming problems on cluster computing systems were investigated. To do this, we used the BSF (Bulk Synchronous Farm) parallel computation model based on the ``master-slave'' paradigm. The BSF-implementation of the NSLP algorithm is described. A scalability upper bound of the BSF-implementation of the NSLP algorithm is obtained. This estimation tells us the following. If all the input data of the problem are changed at each iteration then the upper bound of the BSF-program scalability increases proportionally to the square root of the problem dimension. In this case, the NSLP algorithm implementation in the form of a BSF-program has limited scalability. If each inequality of the constraint system has no more than one coefficient changed during an iteration then the upper bound of the BSF-program scalability increases proportionally to the problem dimension. In this case, the NSLP algorithm implementation in the form of a BSF-program is scalable well. The equations for estimating the parallel efficiency of the BSF-implementation of the NSLP algorithm are also deduced. These equations allow us to conclude the following. If during the iteration all the problem input data are dynamically changed then for the high parallel efficiency it is necessary that the problem dimension is much greater than the square of the number of slaves: $n \gg {P^2}$. However, if in each inequality of the constraint system no more than one coefficient changes during each iteration then for a high parallel efficiency it is necessary that the problem dimension be much greater than the number of slaves: $n \gg P$. The numerical experiments with a synthetic problem showed that the BSF model accurately predicts the upper bound of the scalability of the program that implements the Targeting phase using the BSF skeleton.

As future research directions, we intend to do the following:
\begin{enumerate}[label=\arabic*)]
  \item implement the Qwest phase in C language using the BSF skeleton and MPI-library;
  \item carry out numerical experiments on a cluster computing system using synthetic and real LP problems;
  \item compare the scalability boundaries of the Qwest phase obtained experimentally and analytically.
\end{enumerate}

\end{document}